\documentclass{aastex}
\usepackage{amsmath}
\usepackage{amssymb}

\begin{document}

\title{PAMELA's measurements of geomagnetic cutoff variations during solar energetic particle events}
\shorttitle{PAMELA's measurements of geomagnetic cutoff variations during SEP events}
\shortauthors{Bruno et al.}

%\author{\speaker{Alessandro Bruno}\thanks{A footnote may follow.}\\
%        Department of Physics, University of Bari, Bari, Italy\\
%        E-mail: \email{alessandro.bruno@ba.infn.it}}

\author{
A.~Bruno$^{1,*}$,
O.~Adriani$^{2,3}$,
G.~C.~Barbarino$^{4,5}$,
G.~A.~Bazilevskaya$^{6}$,
R.~Bellotti$^{1,7}$,
M.~Boezio$^{8}$,
E.~A.~Bogomolov$^{9}$,
M.~Bongi$^{2,3}$,
V.~Bonvicini$^{8}$,
S.~Bottai$^{3}$,
U.~Bravar$^{10}$,
F.~Cafagna$^{7}$,
D.~Campana$^{5}$,
R.~Carbone$^{8}$,
P.~Carlson$^{11}$,
M.~Casolino$^{12,13}$,
G.~Castellini$^{14}$,
E.~C.~Christian$^{15}$,
C.~De~Donato$^{12,17}$,
G.~A.~de~Nolfo$^{15}$,
C.~De~Santis$^{12,17}$,
N.~De~Simone$^{12}$,
V.~Di~Felice$^{12,18}$,
V.~Formato$^{8,19}$,
A.~M.~Galper$^{16}$,
A.~V.~Karelin$^{16}$,
S.~V.~Koldashov$^{16}$,
S.~Koldobskiy$^{16}$,
S.~Y.~Krutkov$^{9}$,
A.~N.~Kvashnin$^{6}$,
M.~Lee$^{10}$,
A.~Leonov$^{16}$,
V.~Malakhov$^{16}$,
L.~Marcelli$^{12,17}$,
M.~Martucci$^{17,20}$,
A.~G.~Mayorov$^{16}$,
W.~Menn$^{21}$,
M.~Merg\`e$^{12,17}$,
V.~V.~Mikhailov$^{16}$,
E.~Mocchiutti$^{8}$,
A.~Monaco$^{1,7}$,
N.~Mori$^{2,3}$,
R.~Munini$^{8,19}$,
G.~Osteria$^{5}$,
F.~Palma$^{12,17}$,
B.~Panico$^{5}$,
P.~Papini$^{3}$,
M.~Pearce$^{11}$,
P.~Picozza$^{12,17}$,
M.~Ricci$^{20}$,
S.~B.~Ricciarini$^{3,14}$,
J.~M.~Ryan$^{10}$,
R.~Sarkar$^{22,23}$,
V.~Scotti$^{4,5}$,
M.~Simon$^{21}$,
R.~Sparvoli$^{12,17}$,
P.~Spillantini$^{2,3}$,
S.~Stochaj$^{24}$,
Y.~I.~Stozhkov$^{6}$,
A.~Vacchi$^{8}$,
E.~Vannuccini$^{3}$,
G.~I.~Vasilyev$^{9}$,
S.~A.~Voronov$^{16}$,
Y.~T.~Yurkin$^{16}$,
G.~Zampa$^{8}$,
N.~Zampa$^{8}$,
and V.~G.~Zverev$^{16}$.
}

\affil{$^{1}$ Department of Physics, University of Bari ``Aldo Moro'', I-70126 Bari, Italy.}
\affil{$^{2}$ Department of Physics and Astronomy, University of Florence, I-50019 Sesto Fiorentino, Florence, Italy.}
\affil{$^{3}$ INFN, Sezione di Florence, I-50019 Sesto Fiorentino, Florence, Italy.}
\affil{$^{4}$ Department of Physics, University of Naples ``Federico II'', I-80126 Naples, Italy.}
\affil{$^{5}$ INFN, Sezione di Naples, I-80126 Naples, Italy.}
\affil{$^{6}$ Lebedev Physical Institute, RU-119991 Moscow, Russia.}
\affil{$^{7}$ INFN, Sezione di Bari, I-70126 Bari, Italy.}
\affil{$^{8}$ INFN, Sezione di Trieste, I-34149 Trieste, Italy.}
\affil{$^{9}$ Ioffe Physical Technical Institute, RU-194021 St. Petersburg, Russia.}
\affil{$^{10}$ Space Science Center, University of New Hampshire, Durham, NH, USA.}
\affil{$^{11}$ KTH, Department of Physics, and the Oskar Klein Centre for Cosmoparticle Physics, AlbaNova University Centre, SE-10691 Stockholm, Sweden.}
\affil{$^{12}$ INFN, Sezione di Rome ``Tor Vergata'', I-00133 Rome, Italy.}
\affil{$^{13}$ RIKEN, Advanced Science Institute, Wako-shi, Saitama, Japan.}
\affil{$^{14}$ IFAC, I-50019 Sesto Fiorentino, Florence, Italy.}
\affil{$^{15}$ Heliophysics Division, NASA Goddard Space Flight Ctr, Greenbelt, MD, USA.}
\affil{$^{16}$ National Research Nuclear University MEPhI, RU-115409 Moscow, Russia.}
\affil{$^{17}$ Department of Physics, University of Rome ``Tor Vergata'', I-00133 Rome, Italy.}
\affil{$^{18}$ Agenzia Spaziale Italiana (ASI) Science Data Center, %Via del Politecnico snc,
I-00133 Rome, Italy.}
\affil{$^{19}$ Department of Physics, University of Trieste, I-34147 Trieste, Italy.}
\affil{$^{20}$ INFN, Laboratori Nazionali di Frascati, %Via Enrico Fermi 40,
I-00044 Frascati, Italy.}
\affil{$^{21}$ Department of Physics, Universit\"{a}t Siegen, D-57068 Siegen, Germany.}
\affil{$^{22}$ Indian Centre for Space Physics, 43 Chalantika, %Garia Station Road,
Kolkata 700084, West Bengal, India.}
\affil{$^{23}$ Previously at INFN, Sezione di Trieste, I-34149 Trieste, Italy. }
\affil{$^{24}$ Electrical and Computer Engineering, New Mexico State University, Las Cruces, NM, USA.}

\altaffiltext{*}{Corresponding author. E-mail address: alessandro.bruno@ba.infn.it.}

\begin{abstract}
Data from the PAMELA satellite experiment were used to measure the geomagnetic cutoff for high-energy ($\gtrsim$ 80 MeV) protons during the solar particle events on 2006 December 13 and 14. The variations of the cutoff latitude as a function of rigidity were studied on relatively short timescales, corresponding to single spacecraft orbits (about 94 minutes). Estimated cutoff values were cross-checked with those obtained by means of a trajectory tracing approach based on dynamical empirical modeling of the Earth's magnetosphere. We find significant variations in the cutoff latitude, with a maximum suppression of about 6 deg for $\sim$80 MeV protons during the main phase of the storm. The observed reduction in the geomagnetic shielding and its temporal evolution were compared with the changes in the magnetosphere configuration, investigating the role of IMF, solar wind and geomagnetic (Kp, Dst and Sym-H indexes) variables and their correlation with PAMELA cutoff results.
\end{abstract}

%\FullConference{The 34th International Cosmic Ray Conference,\\
%		30 July- 6 August, 2015\\
%		The Hague, The Netherlands}

\section{Introduction}
Solar Energetic Particle (SEP) events are major space weather phenomena associated with explosive processes occurring in the solar atmosphere, such solar flares and Coronal Mass Ejections (CMEs). SEPs can produce hazardous effects to manned and robotic flight missions in the near-Earth space environment, and 
influence the atmospheric chemistry and dynamics. Large SEP events can strongly perturb the Earth's magnetic field, inducing geomagnetic storms and mo\-di\-fying the Cosmic-Ray (CR) access to the inner magnetosphere. The consequent reduction in the geomagnetic shielding can significantly increase the potential radiation exposure compared with geomagnetically quiet times.
Estimates of geomagnetic cutoffs have been provided by satellite observations and theoretical calculations \citep{Leske,Ogliore,Kress2010} mainly based on tracing particles through models of the Earth's magnetic field \citep{SMART1985,SMART2003}.

In this work we present PAMELA's measurements of the variability of the geomagnetic cutoff during the SEP events on 2006 December 13 and 14,
with the focus on the strong magnetic storm on December 14 and 15.

\section{Data Analysis}

\subsection{The PAMELA experiment}
PAMELA is a space-based experiment designed for a precise measurement of the charged 
cosmic radiation
in the kinetic energy range from some tens of MeV up to several hundreds of GeV \citep{Picozza,PHYSICSREPORTS}.
In particular, PAMELA is providing accurate measurements of SEPs in a wide energy interval \citep{SEP2006,MAY17PAPER}, bridging the low energy data by 
other spacecrafts
and the GLE data by the worldwide network of neutron monitors; in addition, the detector is sensitive to the particle composition and is able to reconstruct flux angular distributions \citep{BRUNOARXIV}, enabling a 
more complete view of SEP events.

The Resurs-DK1 satellite, which hosts the apparatus, was launched into a semi-polar (70 deg inclination) and elliptical (350$\div$610 km altitude) orbit on 2006 June 15. The spacecraft is 3-axis stabilized; its orientation is calculated by an onboard processor with an accuracy better than 1 deg. Particle directions are measured with a high angular resolution ($<$ 2 deg). Details about apparatus performance, proton selection, detector efficiencies and experimental uncertainties can be found elsewhere (e.g. \citep{SOLARMOD}).
The selected data set 
includes protons acquired by PAMELA between 2006 December 12 and 18.

\subsection{Geomagnetic Field Models}\label{Geomagnetic field models}
The analysis described in this work is based on the IGRF-11 \citep{IGRF11} and the TS05 \citep{TS05} models for the description of the internal and external geomagnetic field, respectively. 
The TS05 model is a high resolution dynamical model of the storm-time geomagnetic field, 
based on recent satellite measurements; consistent with the data-set coverage, it is valid
for $X_{GSM}$ $>$ -15 Earth's radii (Re). For comparison purposes, the T96 model \citep{T96} (valid up to 40 Re) was used as well.
Solar Wind (SW) and Interplanetary Magnetic Field (IMF) parameters were obtained from the high resolution (5-min) Omniweb database \citep{OMNIWEB}.

\subsection{Coordinate Systems}\label{Coordinate systems}
Data were analyzed in terms of Altitude Adjusted Corrected GeoMagnetic (AACGM) coordinates, developed to provide a more realistic description of high latitude regions by accounting for the multipolar geomagnetic field. They are defined such that all points along a magnetic field line have the same geomagnetic latitude and longitude, so that they are closely related to invariant magnetic coordinates \citep{Baker, Gustafson, Heres}. The AA\-CGM reference frame coincides with the standard Corrected GeoMagnetic (CGM) coordinate system \citep{BREKKE} at the Earth's surface. Unlike other commonly used variables such as the invariant latitude, 
the computation of such coordinates at low Earth orbits is not significantly affected by the modeling of external geomagnetic sources.

\subsection{Evaluation of geomagnetic cutoff latitudes}
The lowest magnetic latitude to which a CR particle can penetrate the Earth's magnetic field is known as its \emph{cutoff latitude} and is a function of the particle momentum per unit charge, which is referred to as its rigidity. Alternatively one may consider a \emph{cutoff rigidity} corresponding to a given location in space, 
i.e. the minimum rigidity needed to access to that location.
Some complications arise from the presence of the Earth's solid body (together with its atmosphere):
both ``allowed'' and ``forbidden'' bands of CR particle access are present in the so-called ``penumbra'' region \citep{Cooke}.

The numerical algorithm developed to extract cutoff latitudes from the PAMELA data is si\-mi\-lar to one used by \citet{Leske} and \citet{Kress2010}. 
For each rigidity bin, a mean flux was obtained by averaging fluxes above 65 degrees latitude, and the cutoff latitude was evaluated as the latitude where the flux intensity is equal to the half of the average value.

Alternatively, cutoff latitudes were estimated with back-tracing techniques \citep{ALBEDO}. Using the spacecraft ephemeris data, 
and the particle rigidity and direction provided by the PAMELA tracking system, trajectories of all detected protons were reconstructed by means of a tracing program based on numerical integration methods \citep{TJPROG,SMART}, and implementing the afore-mentioned geomagnetic field models.
Trajectories were back propagated from the measurement location 
until they escaped the model magnetosphere boundaries (Solar or Galactic CRs) or they reached an altitude\footnote{Corresponding to the mean production altitude for albedo protons.} of 40 km (re-entrant albedo CRs). 
At a given rigidity, the cutoff latitude was evaluated as the latitude where an equal percentage of interplanetary and albedo CRs was registered.

The calculation was performed for 13 rigidity logarithmic bins, covering the interval 0.39$\div$3.29 GV. Accounting for the limited statistics at highest rigidities, final cutoff values were derived by
fitting averaged
PAMELA observations over single orbital periods ($\sim$94 min).

\begin{figure*}[!t]
\centering
\includegraphics[width=4.5in]{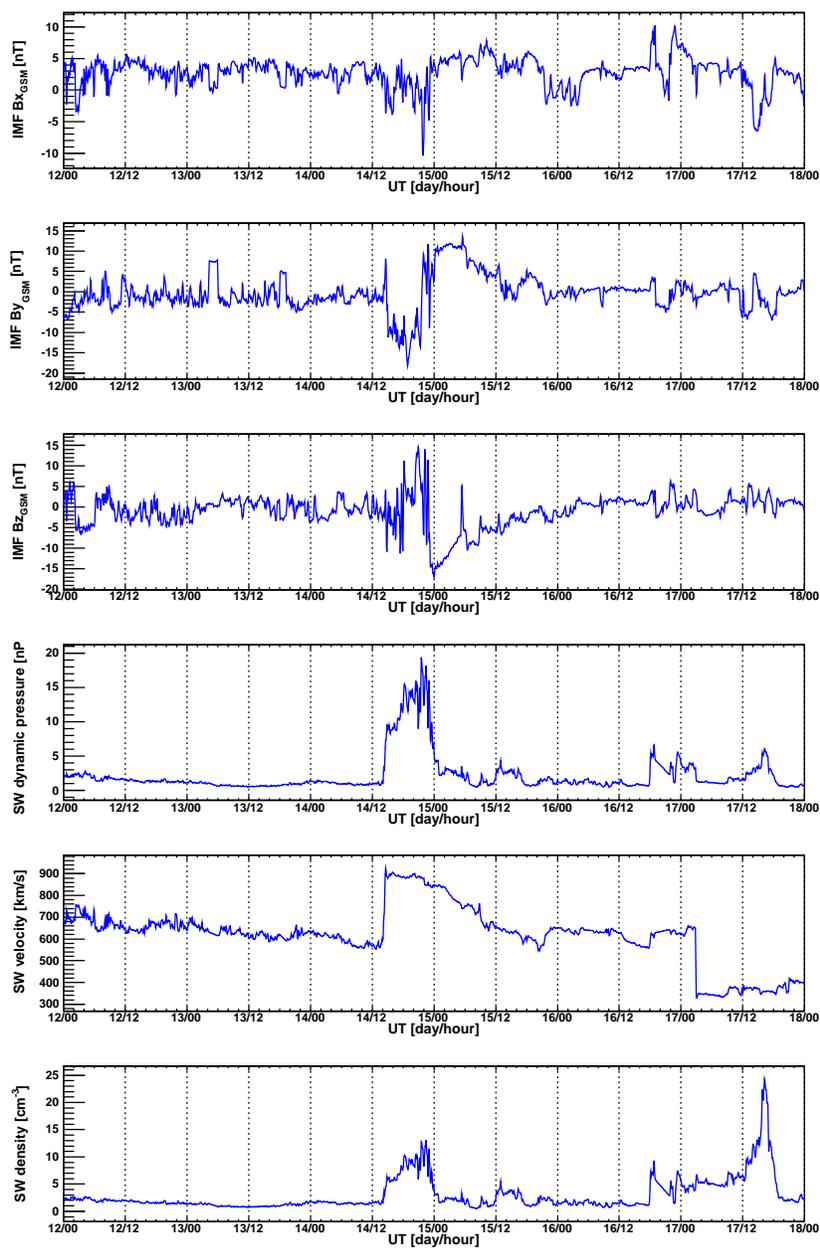}
\caption{Time profiles of the IMF (Bx, By and Bz components in the GSM frame) and solar wind (dynamic pressure, velocity and density) parameters between 2006 December 12$\div$18 \citep{OMNIWEB}.}
\label{corrPlotIMF_PAM_Rsel0_iR20_iAACGL20_k1}
\end{figure*}

\section{The 2006 December 13 and 14 events}
On 2006 December 13 at 02:14 UT, an X3.4/4B solar flare occurred in the active region NOAA 10930 (S06W23; NOAA-STP 2006). This event also produced a full-halo CME with the sky plane projected speed of 1774 km s$^{-1}$. 
The forward shock of the CME reached Earth at about 14:10 UT on December 14, causing a Forbush decrease of Galactic CR intensities that lasted for several days.
Such large events are untypical of the intervals of low solar activity. The flare X1.5 (S06W46) at 21:07 UT on December 14 gave start to a new growth of particle intensity as recorded by PAMELA and other satellites. The maximum energy of protons was below 1 GeV, and therefore no ground level enhancement (GLE) was recorded. The corresponding CME had a velocity of 1042 km s$^{-1}$. 
PAMELA's measurements of the 2006 December SEP fluxes can be found in publications \citep{SEP2006}.

Figure \ref{corrPlotIMF_PAM_Rsel0_iR20_iAACGL20_k1} reports the variations in the
IMF (Bx, By and Bz components in the GSM frame) and
SW (dynamic pressure, velocity and density) variables
between 2006 December 12$\div$18.
The large increase in the SW velocity associated with the leading edge of the CME caused a sudden commencement of a geomagnetic storm.
The initial phase of the storm, lasting up to about 23:00 UT, was characterized by intense fluctuations in the SW density and in all IMF components. 
At a later stage, the IMF Bz component became negative, the SW density decreased, and the main phase of the 
storm started, reaching a maximum between 02:00$\div$08:00 UT on December 15. Another interplanetary shock associated with a different CME was observed on December 16.

\begin{figure*}[!t]
\centering
\includegraphics[width=4.5in]{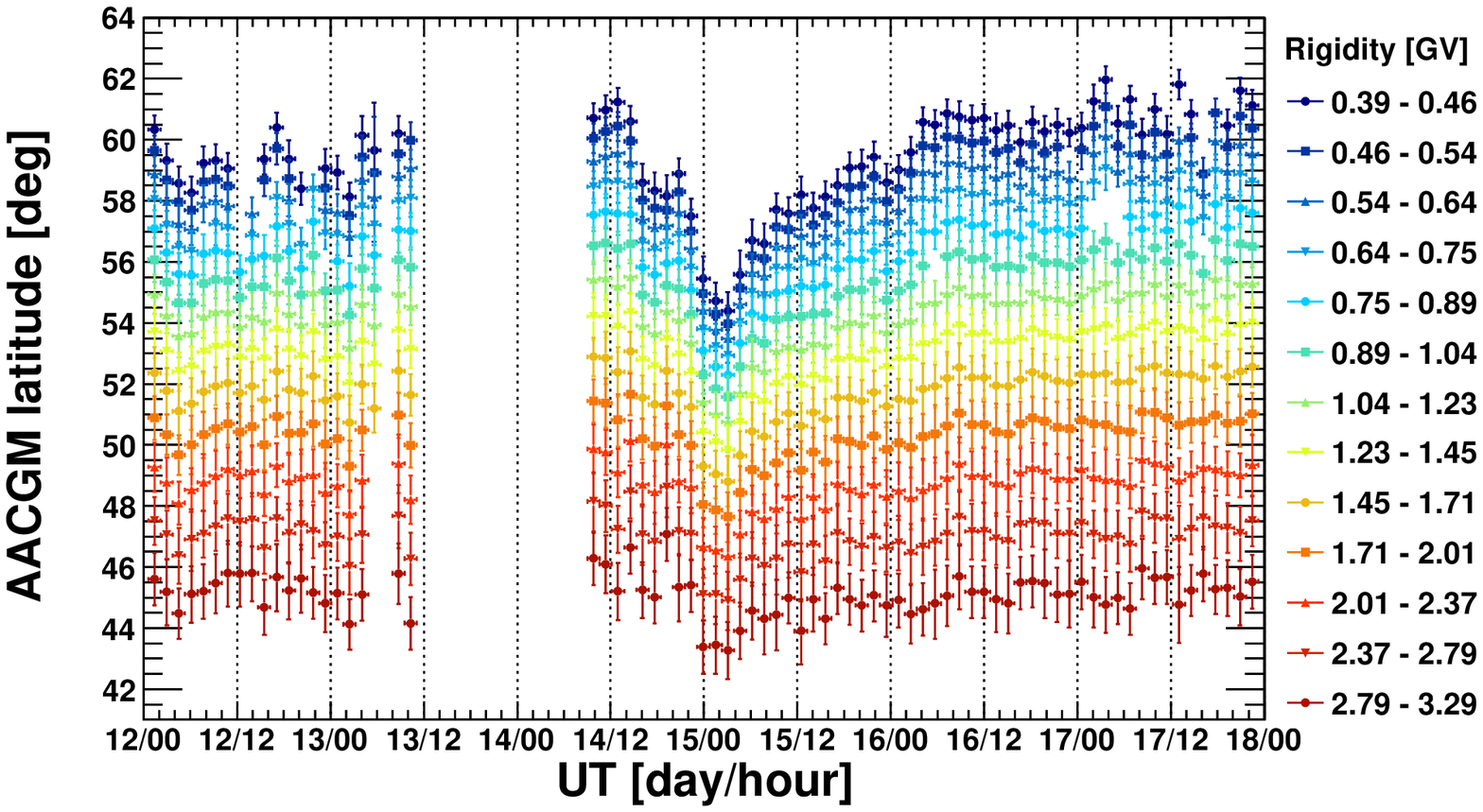}
\caption{Time profile of the geomagnetic cutoff latitudes measured by PAMELA, for different rigidity bins.}% (color code).}
\label{cTimeVarFit_PAM_AACGLat_iR20_iAACGL20_k2}
\end{figure*}

\begin{figure*}[!t]
\centering
\includegraphics[width=4.5in]{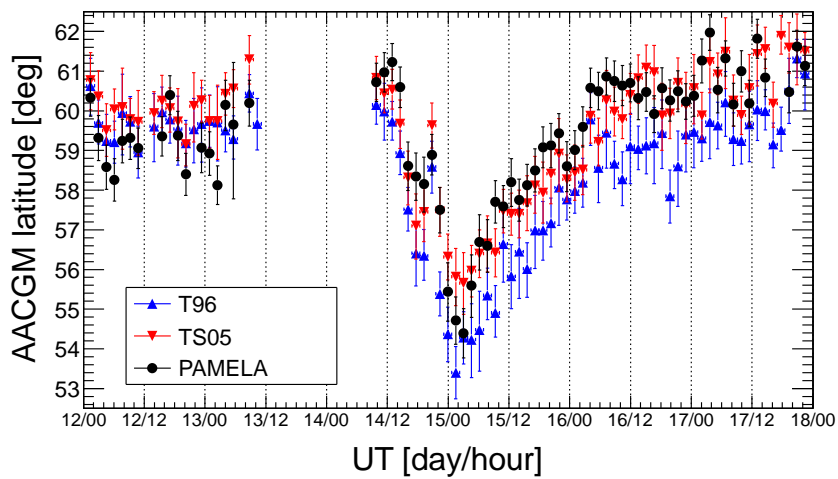}
\caption{Comparison between measured (black) and modeled (blue - T96 model; red - TS05 model) cutoff variations in the lowest rigidity interval: 0.39-0.46 GV.}
\label{comparison}
\end{figure*}

\begin{figure*}[!t]
\centering
\includegraphics[width=4.5in]{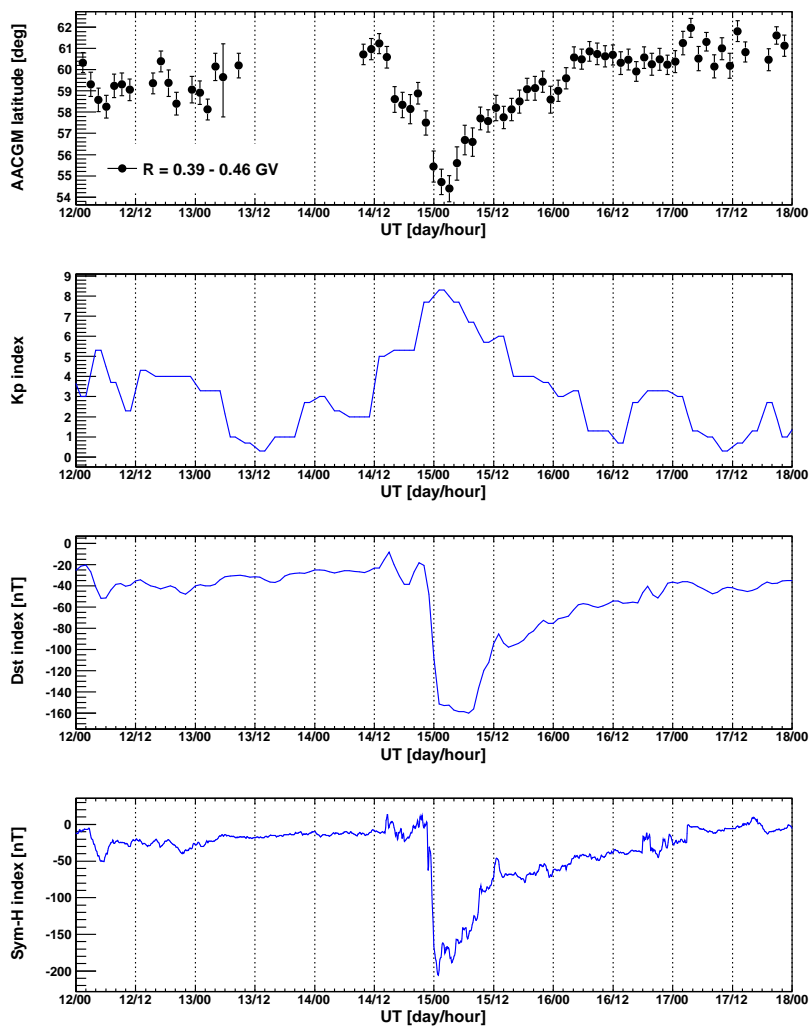}
\caption{Geomagnetic cutoff latitude variations measured by PAMELA in the rigidity interval 0.39-0.46 GV (top panel), compared with the time profiles of Kp, Dst and Sym-H indexes.}
\label{corrPlotGM_PAM_Rsel0_iR20_iAACGL20_k2}
\end{figure*}

\section{Results}
Figure \ref{cTimeVarFit_PAM_AACGLat_iR20_iAACGL20_k2} shows the geomagnetic cutoff latitudes measured by PAMELA for different rigidity bins (color code). Each point denotes the cutoff latitude value averaged over a single spacecraft orbit; the error bars include the statistical uncertainties of the measurement. Data were missed from 10:00 UT on December 13 until 09:14 UT on December 14 because of an onboard system reset of the satellite.
The evolution of the magnetic storm of December 14 and 15 followed the typical scenario in which the 
cutoff latitudes move equatorward as a consequence of a CME impact on the magnetosphere with an associated transition to southward Bz.
The registered cutoff variation decreases with increasing rigidity, with a maximum suppression ranging from about 6 deg in the lowest rigidity bin ($\sim$80 MeV energy) to about 2 deg in the highest rigidity bin ($\sim$3 GeV energy). 

Figure \ref{comparison} reports the comparison between measured and modeled cutoff latitudes, for the lowest rigidity bin 0.39$\div$0.46 GV.
While the T96 model appears to underestimate (up to 4\%) the observations, a much better agreement can be noted between PAMELA and TS05 results. However, the TS05 cutoff latitudes overestimate (up to 2\%) the PAMELA ones during the storm main phase.

Finally, Figure \ref{corrPlotGM_PAM_Rsel0_iR20_iAACGL20_k2} demonstrates the utility of the three indices used to infer cutoff latitude: the magnetic activity index (Kp), the disturbance storm time index (Dst) and the Sym-H index\footnote{Sym-H represents the longitudinally symmetric part of the northward magnetic field variations.}, measured using ground-based magnetometers, at 3-hour, 1-hour, and 1-min resolutions, respectively. In general, the shapes of the time variations in the cutoff measurements are well correlated with corresponding
indexes changes (corresponding correlation coefficients are 0.8, 0.78 and 0.78, respectively). 
A better agreement is observed for Kp during the initial phase of the storm, while the Dst and the Sym-H indexes 
show an improved correlation
during the main and the recovery phases.

\section{Summary and Conclusions}
In this study we have exploited the data of the PAMELA satellite experiment to perform a measurement of the geomagnetic cutoff variations during the long lasting SEP events of 2006 December 13 and 14. A significant reduction in the geomagnetic shielding was observed during the consequent strong magnetic storm on December 14 and 15, with a maximum cutoff latitude suppression of about 6 deg for $\sim$80 MeV protons. Results were compared with those obtained with back-tracing techniques. The observed cutoff variations are well correlated with the time profiles of the geomagnetic indexes (Kp, Dst and Sym-H).

\end{document}